# Graphene and its elemental analogue: A molecular dynamics view of fracture phenomenon

Tawfiqur Rakib, Satyajit Mojumder*, Sourav Das, Sourav Saha, Mohammad Motalab

Department of Mechanical Engineering, Bangladesh University of Engineering and Technology,Dhaka-1000, Bangladesh.

**Abstract**- Graphene and some graphene like two dimensional materials; hexagonal boron nitride (hBN) and silicene have unique mechanical properties which severely limit the suitability of conventional theories used for common brittle and ductile materials to predict the fracture response of these materials. This study revealed the fracture response of graphene, hBN and silicene nanosheets under different tiny crack lengths by molecular dynamics (MD) simulations using LAMMPS. The useful strength of these two dimensional materials are determined by their fracture toughness. Our study shows a comparative analysis of mechanical properties among the elemental analogues of graphene and suggested that hBN can be a good substitute for graphene in terms of mechanical properties. We have also found that the pre-cracked sheets fail in brittle manner and their failure is governed by the strength of the atomic bonds at the crack tip. The MD prediction of fracture toughness shows significant difference with the fracture toughness determined by Griffth's theory of brittle failure which restricts the applicability of Griffith's criterion for these materials in case of nano-cracks. Moreover, the strengths measured in armchair and zigzag directions of nanosheets of these materials implied that the bonds in armchair direction have the stronger capability to resist crack propagation compared to zigzag direction.

Corresponding Author. Tel.: (+880)1737434034; E-mail address: satyajit@me.buet.ac.bd

# 1. INTRODUCTION

Graphene, a single layer carbon atoms arranged in two-dimensional honeycomb structure, has attracted attention owing to its exceptional properties such as- high mechanical strength [1], ultra electric and thermal properties [2, 3]. After the rise of graphene [4], researchers are now looking for other two-dimensional materials similar to graphene. Among those, hexagonal boron nitride and silicene having the same honeycomb structure [5, 6] have allured the interest due to their remarkable mechanical and thermal properties [7, 8]. Due to their excellent properties, these materials are used in nano-electro-mechanical systems (NEMS) and electronic application as electrode, thermoplast and insulator [9-11].

Boron nitride nanotube has the same structure as carbon nanotube in which alternating boron and nitrogen atoms substitute for carbon. hBN also possesses high film strength and dielectric properties similar to graphene and it is also used as a counterpart of graphene in the synthesis of hybrid structure [15]. First principle calculation showed that silicene has similar electronic properties as graphene [12]. Both silicene and graphene are zero band gap semiconductors and their charge carriers are massless Dirac fermions with very high carrier mobilities [13]. So, hBN and silicene are the perfect candidates as elemental analogue of graphene.

Great efforts have been made in characterizing the mechanical properties of graphene [16, 17]. Lee et al. [14] reported that the Young's modulus and intrinsic strength of graphene is 1 ± 0.1 TPa and 130 ± 10 GPa respectively assuming the thickness of graphene as 0.335nm. Ab initio study has been done to evaluate mechanical properties of hBN monolayer [18]. A number of work is done on the fracture behavior of hBN nanotubes [19, 43]. However, the fracture behavior of hBN nanosheet by molecular dynamics simulation is much less explored. Silicene is a material with low fracture strength compared to graphene. Pei et al. [20] studied temperature and strain rate dependent fracture behavior of silicene. They revealed that the Young's modulus and fracture strength of silicene are 82.2 GPa and 12.5 GPa at room temperature.

In material science, fracture toughness is one of the fundamental mechanical properties of any material [21]. The deviation of certain properties in small unit from their bulk counterpart was foreseen by Feynman [22]. Accordingly, nano-sheets of graphene, hBN and silicene have different fracture behavior from the same bulk materials. Yin et al. [23] reported that Griffith criterion is not valid for graphene with cracks smaller than 10nm. However, fracture behavior of hBN and silicene nanosheets with pre-cracks remains unexplored.

It is difficult to conduct a series of controlled experiment by systematically reducing the crack size and retaining all the features of the material in nanoscale. So molecular dynamics simulation is an effective way to explore brittle fracture in atomic scale [24-26]. The main aim of this work is to investigate and compare fracture phenomenon of graphene, hBN and silicene with various length of cracks under uniaxial tension using molecular dynamics simulation. In this study, fracture strength of these materials have been compared both in armchair and zigzag directions. The uniqueness of mechanical properties of these materials have been explained with the help of simulation results and failure patterns.

## 2. SIMULATION METHOD

All calculations were performed using LAMMPS (Large-scale Atomic/Molecular Massively Parallel Simulator) software package [29] and the simulation systems were constructed to meet a periodic boundary conditions. The equation of atomic motion was integrated with time step 1 fs. The geometries were relaxed by conjugant gradient minimization scheme. Then the system is equilibrated at NVE ensemble for 100 ps. It is again equilibrated by using isothermal-isobaric (NPT) simulations for $10^5$ time steps at specified temperature. Then a uniaxial stress is applied along armchair and zigzag direction at a constant strain rate $10^9$ s$^{-1}$. Each deformation simulations were performed at low temperature 100 K because of the stability of material properties and remarkable importance of low temperature in electronics [27]. Under deformation, the atomic stress of each atom in the simulation was calculated according to the equation [28]:

$$\sigma_{ij}^{\alpha} = \frac{1}{\Omega^{\alpha}} \left( \frac{1}{2} m^{\alpha} v_i^{\alpha} v_j^{\alpha} + \sum_{\beta=1,n} r_{\alpha\beta}^{j} f_{\alpha\beta}^{i} \right) \qquad (1)$$

Where i and j denote indices in the Cartesian coordinate system; α and β are the atomic indices; $m^\alpha$ and $v^\alpha$ denote the mass and velocity of atom α; $r_{\alpha\beta}$ is the distance between atoms α and β; $f_{\alpha\beta}$ is the force between atoms α and β; $\Omega^\alpha$ is the atomic volume of atom α. The stress during deformation was obtained by averaging the stress of each atom of the system.

For different materials, the atomic interactions in the simulations were defined by different potentials. For graphene, Adaptive Intermolecular Reactive Empirical Bond Order (AIREBO) potential was employed which is widely used to study the mechanical properties of carbon materials e.g. Fullerene [30], CNT [31] and graphene [23, 25]. In LAAMPS, the default C-C bond cut off distance is 1.7. But it is obtained from the literature that this value highly deviates from the real value of C-C bond cut off distance. To avoid non-physical strengthening before fracture, cut-off distance is choosen $r_c = 2.0$ for C-C bonds in graphyne and polycrystalline graphene [32-34]. He et al. [35] simulated graphene nanosheets under tension at different cut off distance and reported that there is no big difference between simulation and experimental results when the cut off distance is in the range of 1.92-2.00 Å. They choose cut off distance 1.95 for their simulations. To facilitate our simulation, we adopt $r_c = 1.95$ Å beyond which C-C bond breaks. AIREBO potential represents sum of all interactions among the atoms including LJ term, covalent bonding and torsional interactions. AIREBO improves Brenner's REBO potential [37] considering all the interactions and predicts the result with more accuracy. In the case of graphene, the bond length of carbon is taken 0.142 nm and the thickness of the sheet is 0.34 nm.

In the MD simulations of silicene, an optimized Stillinger-Weber (SW) potential was implemented to describe the atomic interactions of silicon atoms [38]. Our simulation results show that this atomic potential maintains the initial buckled structure of silicene after relaxation. If the Tersoff potential is used for silicon, a flat structure is obtained after relaxation instead of the original buckled structure [39]. This implies the inapplicability of Tersoff potential. MEAM potential also gives an initial buckling structure in silicene which gives the buckling height 0.85 Å [20]. In our study, we obtain the buckling height 0.44 Å which is reasonably in good agreement with first principle calculations [40]. Since low buckling height is an important quality of silicene, the optimized SW potential used in our simulation predicts the atomic interactions of silicon atoms more accurately. For silicene, Si-Si bond length is 0.228 nm and thickness of the sheet 0.313 nm.

In the simulation method for hBN, the B-N bond length and thickness of the sheet is taken to be 0.145 and 0.33 nm respectively. Tersoff potential [36] was adopted to describe the interatomic interactions of Boron and Nitrogen atom due to its success in studies done by researchers [41].

In our study, uniaxial stress along armchair and zigzag directions is applied in graphene, hBN and silicene nanosheet of approximately 30 nm×30 nm size. An optimum size is taken to reduce sample size effect on both fracture mode and load without significantly rising computational cost. We depicted the stress and strain variations at different crack length of 1, 1.5, 2, 2.5, 3 nm approximately. The deviation of properties in nano-level from the bulk phenomenon is shown by the calculation of fracture toughness from the MD simulations and Griffith criterion. The pre-cracks in the material are formed by removing a line of atoms in a direction perpendicular to loading direction. The difficulties of maintaining the length of the crack and similar sharpness in each crack enables removal of a line of atom as the best possible criteria to form pre-crack in the nanosheets.

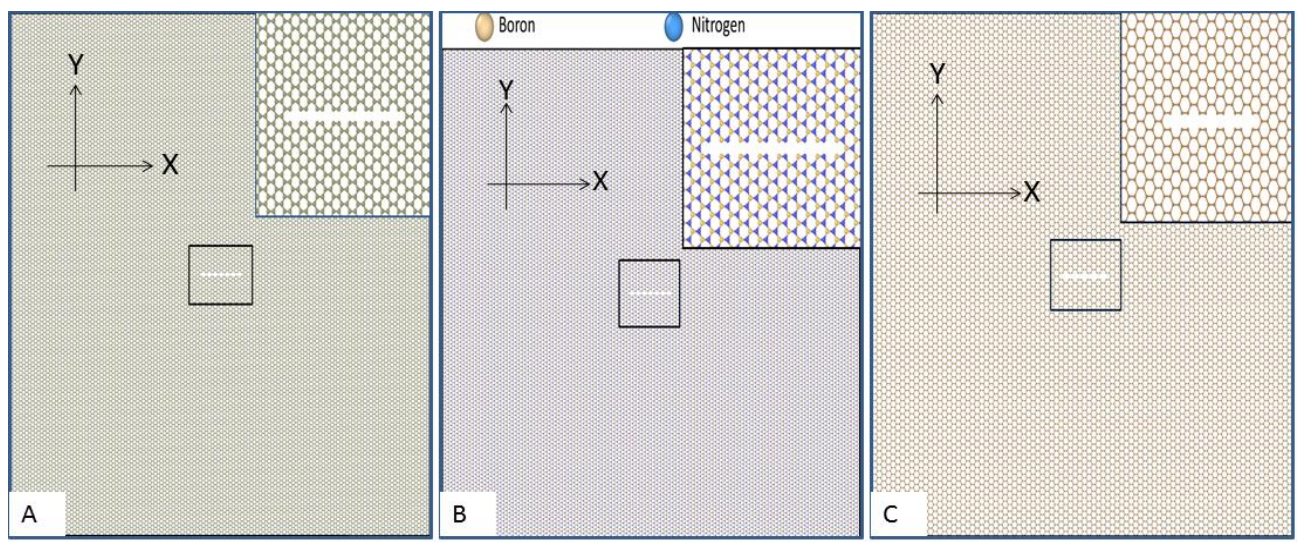

Figure 1: Illustration of simulation model for A) Graphene B) Hexagonal Boron Nitride and C) Silicene with their zoomed view of 3 nm crack. The sheet dimension is 30 nm × 30 nm approximately. Here, the armchair and zigzag edges are shown in X and Y directions respectively. In Figure 1(B), boron and nitrogen atoms are shown by golden and blue color respectively.

# 3. RESULTS AND DISCUSSION

## 3.1 Method Validation

In order to validate our approach, Young's modulus of graphene, hBN, and silicene nanosheet materials have been calculated. For graphene, a 40 Å × 40 Å sheet has been taken in which uniaxial stress is applied in armchair direction at 300K. The AIREBO potential has been used for graphene and an optimized SW potential has been used for silicene. Most of the studies implemented MEAM potential for silicene. We have used an optimized SW potential because low buckling height = 0.44 Å is obtained which is in par with the first principle study. On the other hand, to validate Young's modulus of silicene, uniaxial stress is applied in armchair direction of a 33 nm × 33 nm sheet at 300K temperature.

For hBN, since the literature only have Young's modulus values for nanotube and nanoshells, a model of Boron Nitride nanotube of outer diameter = 40 nm and length = $4 \times 10^3$ nm has been developed to find its Young's modulus at 300K. Periodic boundary conditions and distance controlled tensile loading have been applied in the axial direction of the nanotube and the other two directions have been set as free. In this case, Tersoff potential has been used. All the data obtained from the simulations and literatures are listed in table 1 which show a good agreement among them. This part of the study also reveals the reliability of the potential files used for respective materials.

Table 1: Comparison of Young's modulus with literature data

| Material | Methods | Young's Modulus (GPa) |
|---|---|---|
| Graphene | MD (Nanosheet) | 1050 [42] |
| | Experimental | 1000±100 [14] |
| | MD (Nanosheet) | 989.2 [Present work] |
| hBN | MD (Nanotube) | 505-1031 [43] |
| | MD (Nanotube) | 771.4 [Present work] |
| Silicene | MD (Nanosheet) | 82.2 [20] |
| | MD (Nanosheet) | 85.3 [Present work] |

## 3.2 Stress-strain Behaviour in Pristine Graphene and its Elemental Analogues

The simulated stress- strain curves of Graphene, hBN and silicene in their pristine form for tensile loading in armchair and zigzag direction is shown in figure 2. From the analysis, it can be said that Graphene is the strongest material among all with Young's modulus 961 GPa and fracture strength 123 GPa and 107 GPa in case of armchair and zigzag directions (Bond breaking directions) respectively. These values are relevant with previously reported values [42, 14]. Graphene also exhibits a good amount of failure strain which corresponds to its stretchy behavior [44]. So, Graphene is more flexible to deform and harder to break. In case of hexagonal boron nitride, it is also prominent to a large breaking stress and strain closer to that of graphene. It has been reported that the tensile strength of hBN is very close to graphene for the same loading rates [41]. The Young's modulus of hBN has been obtained as 771 GPa and 768 GPa respectively for armchair and zigzag bond breaking directions and these values are very close to graphene. It is seen that silicene suffers high strain while the fracture stress is far below than that of hBN and graphene. For silicene, the irreversible changes that happen during tensile loading is more distinguishable than the other two [45]. The presence of buckling in silicene also enables its flexibility in deformation at low stress. The Young's modulus of silicene has been found as 88.1 GPa (Armchair) and 87 GPa (Zigzag) which is quite lower than bulk silicon (170 GPa) [46]. But the fracture stress of silicene is 30.6 GPa (Armchair) and 27.8 GPa (Zigzag) which is much higher than that of bulk silicon (7 GPa) [47]. So compared to silicon, silicene is an effective stretchy material with high breaking stress.

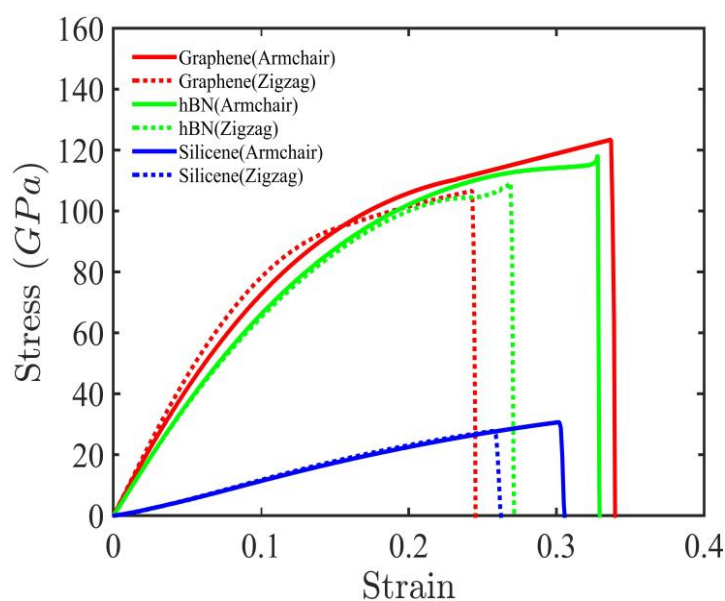

Figure 2: Stress-strain curves for graphene, hBN and silicene in armchair and zigzag bond breaking directions under uniaxial tension at 100 K and loading rate $10^9$ s$^{-1}$. The non-linear responses are evident in the stress-strain curves and graphene shows more non-linear response than any other [14, 48]

## 3.3 Stress-strain Curves of Pre-cracked Graphene, hBN and Silicnene

The stress-strain relationship for pre-cracked graphene, hBN and silicene under uniaxial loading along the direction perpendicular to crack are shown in Figure 3. The pre-cracks are formed by removing a line of atom from either armchair or zigzag direction and loading is applied perpendicular to it. Both the fracture stress and strain are smaller than the pristine form due to the presence of pre-cracks. The pre-cracks cause rupture at a lower value of strain initiating bond breaking near the cracks. The local stress near the crack gets too much high and bond breaking starts near the cracks (Clear evidence shown in Figure 7). It was previously reported that the failure in this type of material is brittle in nature [23, 41]. Our stress-strain diagram also predicts a brittle failure showing no difference between ultimate stress and breaking stress. So, once a bond is broken near the crack, the failure becomes imminent. It is evident that graphene remains the strongest material even in the presence of cracks though the fracture stress and strain are reduced considerably. That means the C-C bonds at the crack tip in graphene resist failure more than that of B-N and Si-Si bonds. Due to the formation of cracks, the strength of silicene is reduced by 50% when compared to the strength of pristine sheet. That's because the Si-Si bonds at the crack tip become very weak due to buckling of silicene structure after relaxation. It is also worth mentioning that the nonlinearity of stress-strain relations in each material remain unchanged in presence of crack. The direction of a crack in a material also predicts the direction of crack propagation in these materials [23, 25].

In addition, it can be seen in figure 3 that the fracture stress and strain in the armchair direction are always slightly higher than that in the zigzag direction. This slight difference can easily be explained by analysing the structure of the nanosheets which are almost similar in case of graphene, hBN and silicene. In the armchair direction there are six stress bearing bonds per unit cell among which two bonds are parallel to the armchair direction and the remaining four have relative angles of $30^0$ with the armchair direction. However, in zigzag direction the number of bonds bearing load are four per unit cell. These bonds make an angle of $60^0$ with the zigzag direction. So, the extra parallel bonds in the armchair direction are the cause of more resistance to fracture than that in the zigzag direction. So, assuming all conditions ideal, the mechanical properties in the armchair direction are always slightly better than that of zigzag direction in all these 2D materials of same type.

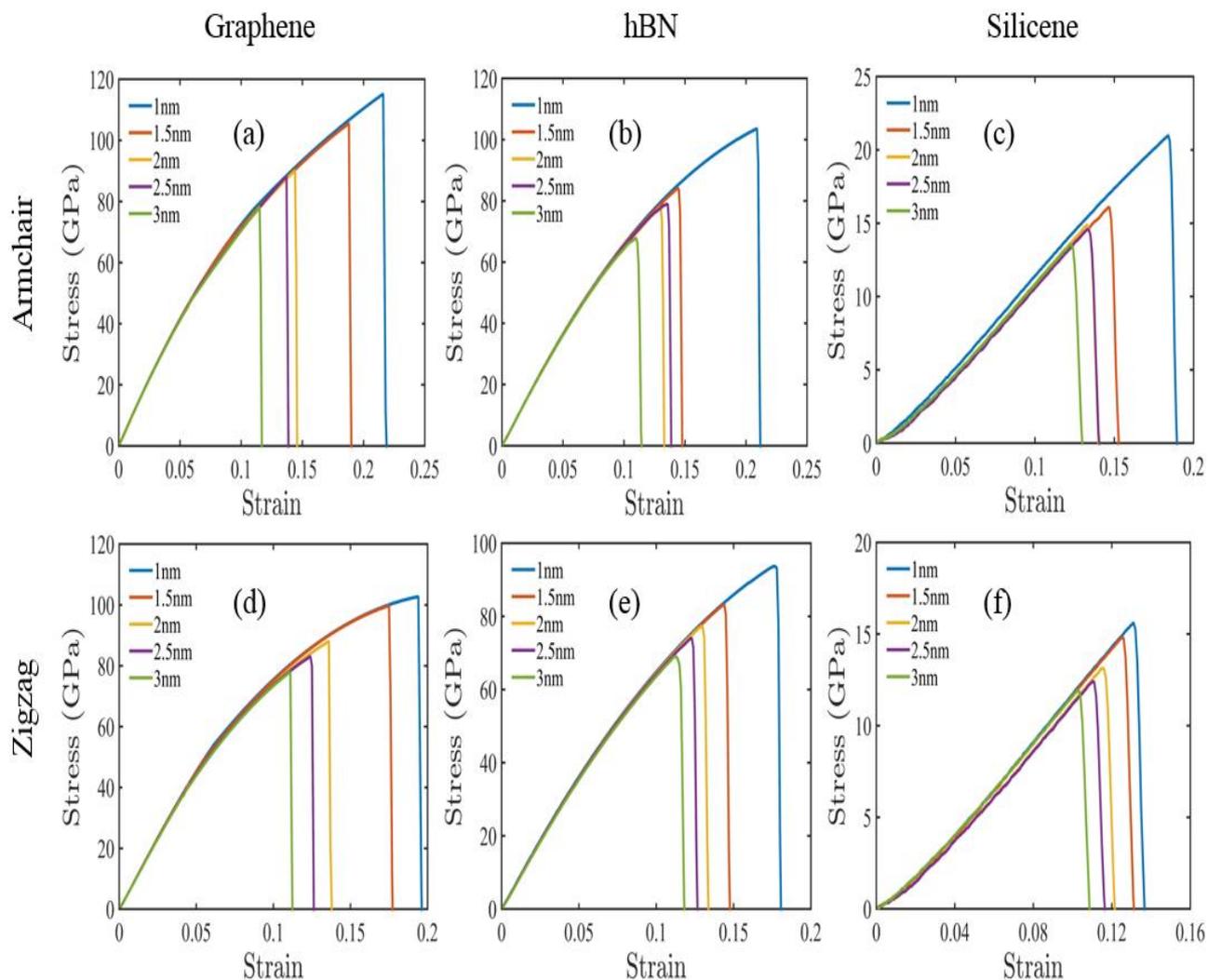

Figure 3: Stress-strain curves for different precracked sheet of (a), (d) for graphene, (b), (e) for hBN and (c), (f) for silicene with crack lengths 2a = 1, 1.5, 2, 2.5, 3 nm. The bond breaking takes place in armchair and zigzag direction under uniaxial tension perpendicular to the direction of bond breaking.

## 3.4 Variation of Fracture Stresses and Strains with Crack Lengths

In this study, we simulated pre-cracked sheets of graphene, silicene and hBN at different crack length under uniaxial tension along the direction perpendicular to crack direction. The variation of fracture stresses and strains due to different crack lengths are presented in figure 5. It is observed that both the fracture stress and strain decreases with the increase of crack length. It indicates that faster bond breaking is initiated by larger cracks by forming a plastic zone at the crack tip. This plastic zone exhibits an irreversible deformation due to the high local stress at the tip and consequently promotes bond breaking. But in case of hexagonal boron nitride, a different phenomenon is noticed for 2.5 nm crack length. In this particular case, both fracture stress and strain for crack length 2.5 nm is slightly more than that of 2 nm crack length. The cause is attributed to the bluntness of the crack tip. An asymmetrical blunt crack is formed to maintain 2.5 nm crack length at the centre of the sheet which gives rise to this anomaly (See figure 4). It was previously reported that the failure of material is delayed by the radius of the crack tip [49].

It is stated that the bonds in armchair direction resist fracture better than those in zigzag direction. Figure 5 also validates this statement for precracked sheet of graphene, hBN and silicene. In case of cracked sheets, the bonds in the armchair direction for silicene strain more than hexagonal boron nitride except in case of 2a = 1nm. Such straining nature of silicene is caused by those two parallel bonds in armchair direction of the unit cell of silicene; which are more stretchy in nature and flexible in deformation. From the figure 5, it is also evident that silicene fails at lower value of fracture stress than hBN although the strains are similar to hBN. It proves that silicene is a unique material which deforms easily at low stress [50]. The decrement of fracture stress is considerably less in silicene than that of the other two materials with the increase of crack length. This is because the local stress concentration remains unaffected with the increase of crack length. The slight decrease in fracture stress is attributed to diminution of bonds due to removal of atoms when forming pre-cracks.

### 3.5 Theoretical and Molecular Dynamics Predictions of Fracture Toughness

The theoretical prediction of fracture toughness is made by Griffith theory of brittle fracture. According to Griffith, the critical fracture stress of a stripe with central crack is expressed by [21] –

$$\sigma_f = \frac{1}{f(\Omega)} \sqrt{\frac{E\Gamma}{\pi a}} \tag{2}$$

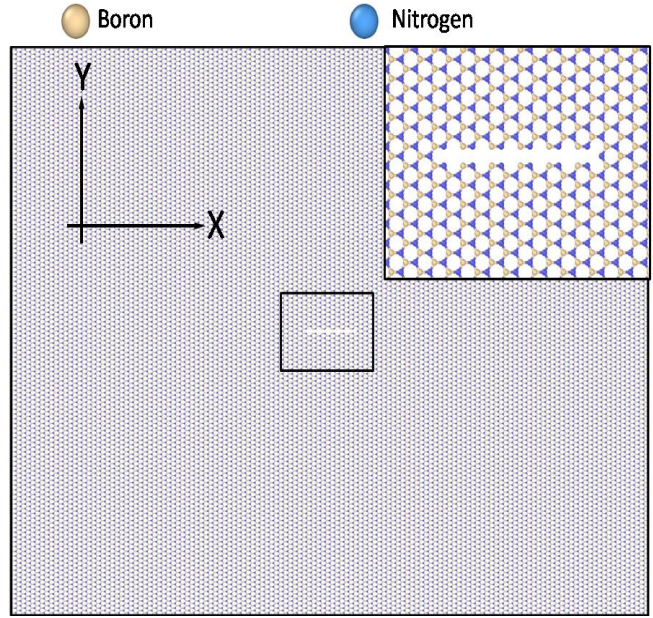

Figure 4: Illustraion of hBN model for 2.5 nm crack length with the zoomed view of crack. The crack tips at the right side is blunter than the left.

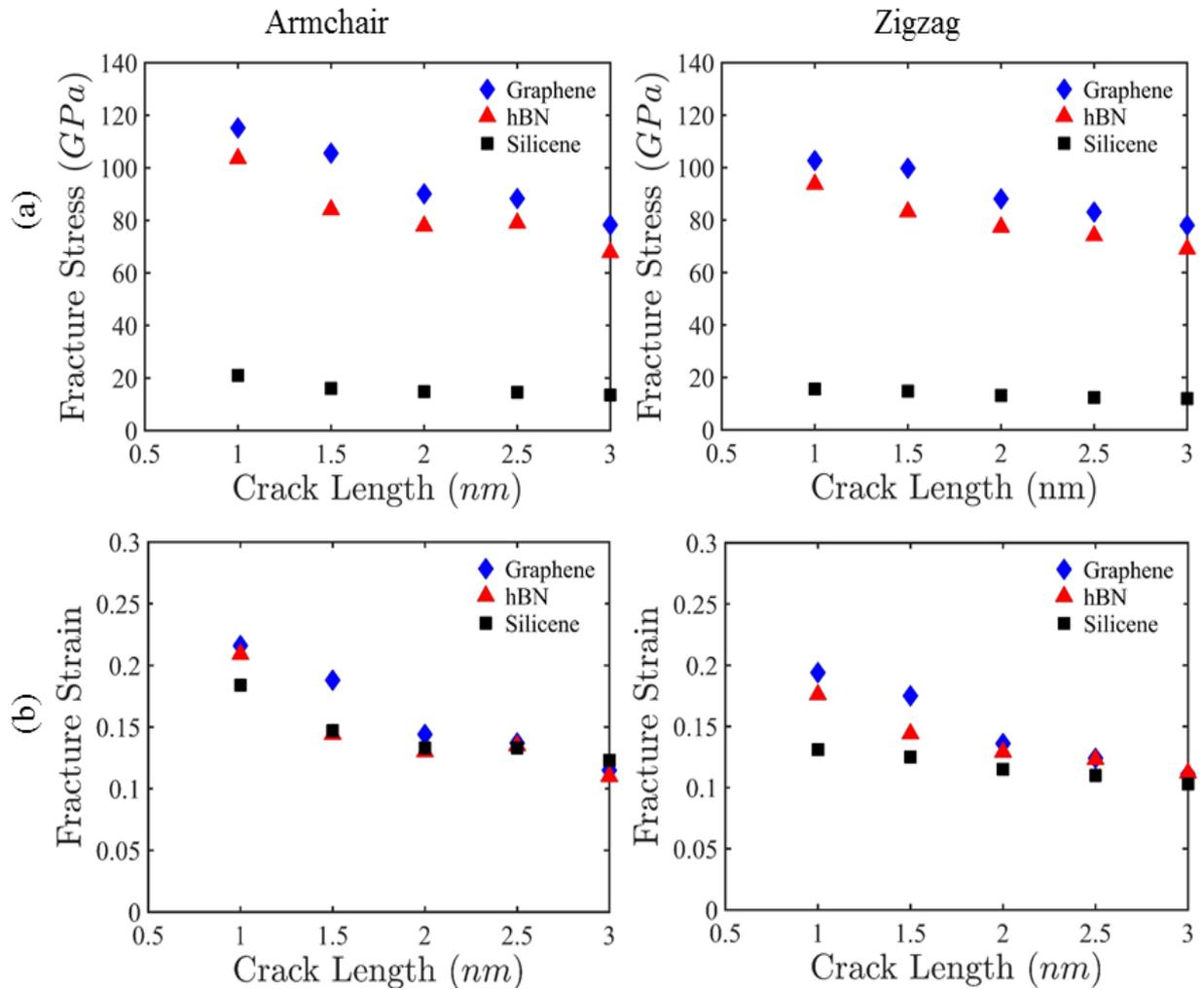

Figure 5: Variation of (a) Fracture stress and (b) Fracture strain at different crack lengths 2a = 1, 1.5, 2, 2.5, 3 nm for Graphene, hBN and silicene in case of armchair and zigzag directional bond breaking under uniaxial tension. Here, the temperature is maintained at 100 K and loading rate is $10^9$ s$^{-1}$.

Where, E is the Young's modulus, Γ is the surface energy for 3D material and edge energy for 2D material and 'a' is half the crack length. The function f(Ω) is a geometrical factor given by-

$$f(\Omega) = [1 - 0.025\Omega^2 + 0.06\Omega^4] \times \text{Sec}\left(\frac{\pi\Omega}{2}\right)^{1/2} \tag{3}$$

Here, Ω = 2a/W; W is the width of the strip with a central crack of length 2a. The edge energy Γ is obtained by calculating the difference between energy of the samples with and without edges. Since edge energy is considered as constant for a specific crack orientation irrespective of the crack length, we assumed an average value of Γ for all crack lengths. The value of Γ for graphene is 10.0 Jm$^{-2}$ [26] by first principle calculation and 11.8 Jm$^{-2}$ [51] by MD simulation. The value of Γ obtained for graphene in our MD simulation is 11.79 Jm$^{-2}$ in armchair direction. So, these values are in close agreement with the previous studies. This validates our approach for calculating edge energy of all the materials under study. By calculating the edge energy, the fracture toughness is determined by the following equation derived from eqn (2)-

$$\sigma_f \sqrt{a} = \frac{1}{f(\Omega)} \frac{E\Gamma}{\pi} \tag{4}$$

On the other hand, the molecular dynamics prediction of fracture toughness is calculated by $\sigma_f\sqrt{a}$ where $\sigma_f$ is the fracture stress of the pre-cracked sheet. Fracture toughness of graphene calculated in this way has its discrepancy with the experimental value 2.25 MPa√m [51]. This discrepancy arises owing to the crack bluntening, crack orientation, polycrystalline structure and lattice trapping.

Fracture toughness is one of the important material property which measures the resistance of fracture of a material containing crack. From the graphs, it is understood that graphene has the highest ability to resist fracture in its pre-cracked sheet and silicene has the lowest. Fracture toughness of hBN also agrees with the fact that hexagonal Boron Nitride is a material with strength closer to graphene.

From figure 6, it is seen that the prediction of fracture toughness by MD simulation deviates from Griffith theory. The reason behind this is that Griffith theory was formulated based on a stress analysis of bulk material. But our study is based on the nanosheets of graphene, hBN and silicene. MD results of samples with large cracks match well with the theoretical prediction but the magnitude of fracture toughness in samples with smaller crack shows large deviation [23]. In our study, we simulated graphene, hBN and silicene with crack lengths upto 3 nm. Therefore, the results of MD analysis in figure 6 show significant transgression not only in graphene but also in its elemental analogues. On the other hand, due to hexagonal honeycomb structure, it is impossible to maintain the pattern of Griffith crack in the nanosheet of these materials. It is also mentionable that Griffith strength for nanosized crack is governed by the local properties of bond breaking at the crack tip instead of the global energy balance.

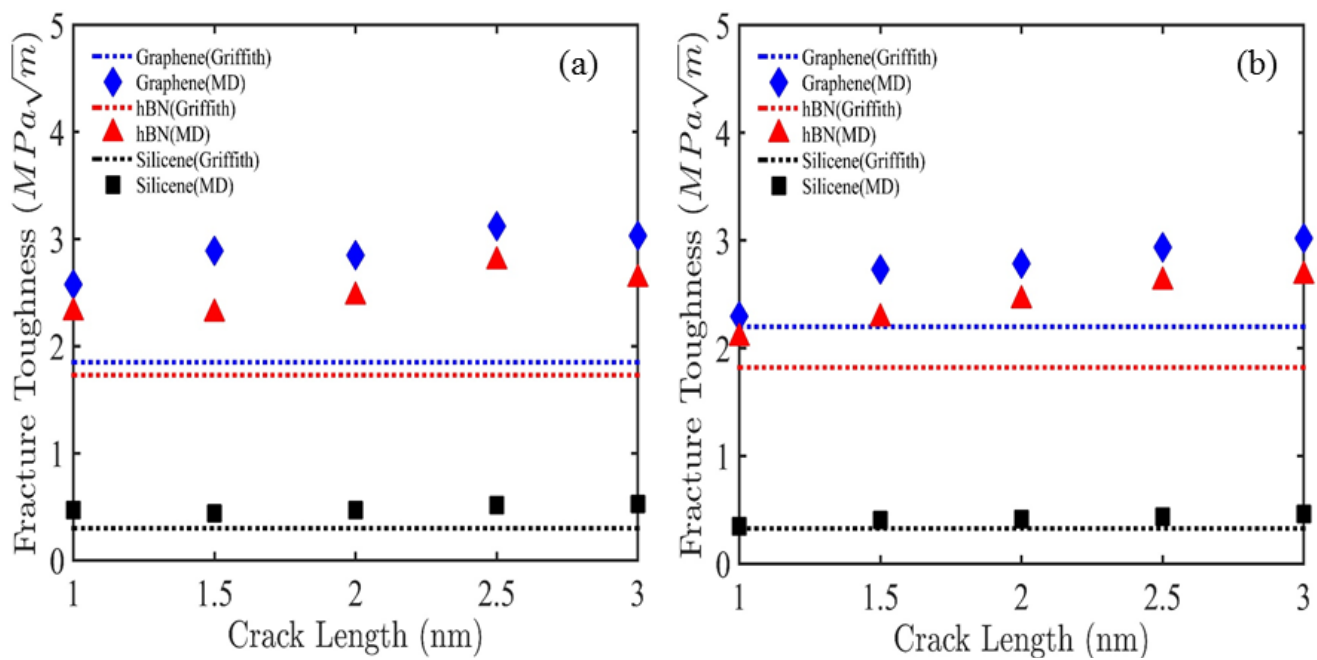

Figure 6: Prediction of fracture toughness by Griffith theory and MD simulation at different crack length along (a) Armchair and (b) Zigzag directional bond breaking. Fracture toughness is constant in case of Griffith's theoretical prediction. Because an average value of edge energy $\Gamma$ for all crack length is taken. So, the theoretical fracture toughness is an average prediction.

It is also seen from figure 7 that the crack propagation in case of armchair direction always tries to kink towards zigzag direction. Deviation of actual crack from the initial crack plane requires a high strain energy release to extend the crack. This type of crack propagation is contrary to the popular assumption that cracks propagate along the crack edges which is also energetically favorable. So, our MD results of fracture toughness in armchair direction shows considerable discrepancy than theoretical results.

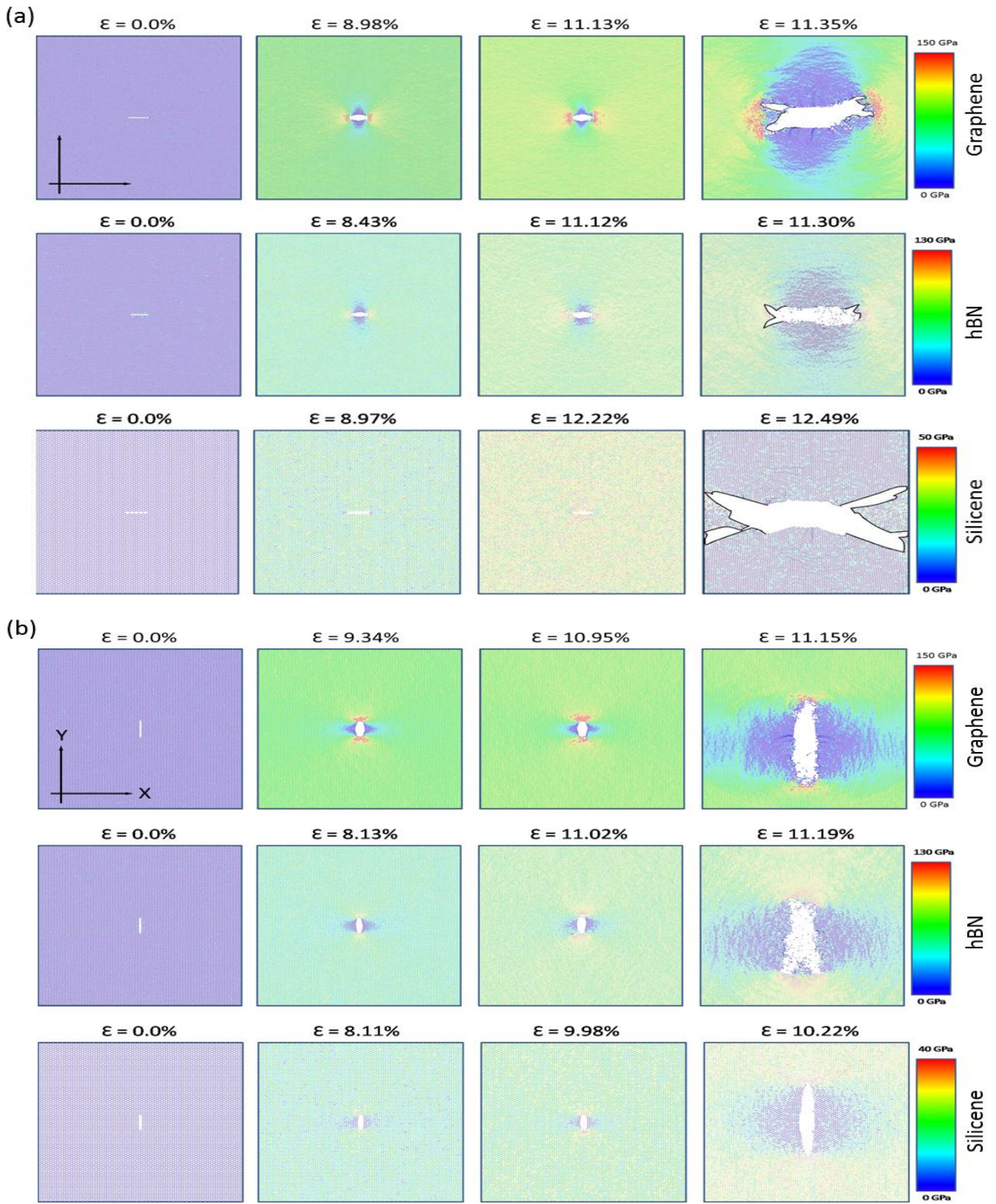

Figure 7: The stress distribution, deformation process and crack propagation of graphene, hBN and silicene in (a) Armchair and (b) Zigzag directional bond breaking. The color bar shows the stress in GPa. Here, X and Y axis represents armchair and zigzag edge respectively. The loading rate of $10^9$ s$^{-1}$ is applied along (a) zigzag and (b) armchair directions repectively. The black lines in Figure (a) points out branching (indicates brittleness of the materials) and kinking towards zigzag direction for all materials. The red dumb-bell shaped zone near the crack tip is the plastic zone where irreversible deformation takes place. The area of this zone increases as the strain increases initiating fracture in the material.

The deviation of fracture toughness by molecular dynamics simulations for silicene has very small difference with theoretical prediction. This is due to the fact that the fracture stress of silicene decreases slightly even if the crack length increases. The local stress concentration at the crack tip do not change considerably with the increase of crack length. This slight decrease of fracture stress is only attributed to the presence of less number of bonds as a result of removal of more atom for larger crack length.

From figure 6, our MD results also suggest that the energies required to tear apart the bonds in armchair direction is larger than that needed to tear apart the bonds in zigzag direction. This result implies with the stress-strain analysis.

## 4. CONCLUSIONS

From our investigation, we find that the stresses at fracture for graphene, hBN and silicene are dependent on loading direction, bond strength, crack length, and crack tip sharpness. The key points of our findings are listed below:

☐ Graphene is the strongest among the three materials. Hexagonal boron nitride shows mechanical properties that are close to graphene. Silicene has been found as the weakest among these three materials. However, it shows larger deformation under low stress conditions than the other two materials.

☐ All these materials exhibit brittle type of failure behavior.

☐ Bonds in the armchair direction show greater tendency to resist fracture than in zigzag direction for all the materials.

☐ Fracture stress and strain decrease with increasing crack length. However, the fracture stresses have been observed to increase due to the reduced crack tip sharpness (greater bluntness) when larger crack lengths are considered. So the sharpness of the crack tip governs the failure of the pre-cracked sheets of these materials.

☐ Fracture toughnesses of graphene, hBN and silicene predicted by MD simulation show significant deviations from that predicted by Griffith theory restricting the applicability of Griffith model in case of nano-cracks. The deviation has been observed higher in armchair direction and it is the highest (about 40%) for graphene. This is due to the fact that the crack propagations in armchair direction in these nanosheets always tend to kink towards the zigzag direction. This deviation of crack path requires high strain energy which accounts for the highest deviation of fracture toughness by MD simulation from Griffith model in case of armchair direction for all the three materials.

## 5. ACKNOWLEDGEMENT

The authors would like to thank Department of Mechanical Engineering, BUET for their continuous support of simulation facilities.